\documentclass[conference]{IEEEtran}
\IEEEoverridecommandlockouts
\usepackage{amsfonts}
\usepackage{enumerate,fancybox,url,ascmac,here,comment,wrapfig,cite}
\usepackage{amsmath,amssymb,amsfonts,bm}
\usepackage{algorithmic,algorithm}
\usepackage{graphicx}
\usepackage{textcomp}
\usepackage{mathtools}
\usepackage{xcolor}
\usepackage{ascmac}
\usepackage{physics}
\usepackage{tikz}
\usepackage{mathrsfs}

\newtheorem{algo}{Algorithm}
\newtheorem{rem}{Remark}
\newtheorem{theo}{Theorem}
\newtheorem{lem}{Lemma}
\newtheorem{coro}{Corollary}
\newtheorem{defi}{Definition}
\newtheorem{exam}{Example}

\def\BibTeX{{\rm B\kern-.05em{\sc i\kern-.025em b}\kern-.08em
    T\kern-.1667em\lower.7ex\hbox{E}\kern-.125emX}}

\newcommand{\setint}[1]{[\hspace{-0.5mm}[#1]\hspace{-0.5mm}]}
\newcommand{\1}{\mbox{1}\hspace{-0.25em}\mbox{l}}
\newcommand{\mat}[1]{\mathbf{#1}}
\newcommand{\mtt}[1]{\mathtt{#1}}
\newcommand{\mbb}[1]{\mathbb{#1}}
\newcommand{\Pure}[2]{P \bigl(\mathcal{H}_{#1}^{\otimes {#2}} \bigr)}
\newcommand{\State}[2]{S \bigl(\mathcal{H}_{#1}^{\otimes {#2}} \bigr)}
\makeatletter\renewcommand{\ALG@name}{Subroutine}\makeatletter
\newcommand{\digrV}[0]{\mtt{V}(\mtt{G}_{\bm{x}, \bm{y}})}
\newcommand{\digrA}[0]{\mtt{A}(\mtt{G}_{\bm{x}, \bm{y}})}
\newcommand{\Pset}[0]{\mtt{P}_{\bm{x}, \bm{y}}}
\newcommand{\Ptop}[0]{\mtt{P}_{\top}}
\newcommand{\Pbot}[0]{\mtt{P}_{\bot}}

\newcommand{\cha}[1]{\textcolor{black}{#1}}
\newcommand{\M}{\cha{\bm{m}}}

\begin{document}

\title{Decoding Algorithm Correcting Single-Insertion Plus Single-Deletion for Non-binary Quantum Codes}

\author{\IEEEauthorblockN{Ken Nakamura and Takayuki Nozaki}
\IEEEauthorblockA{
Dept.\ of Informatics, Yamaguchi University, JAPAN \\
Email: {\tt \{d027vbu, tnozaki\}@yamaguchi-u.ac.jp}}
}

\maketitle

\begin{abstract}
  In this paper,
  we assume an error such that 
  a single insertion occurs and then a single deletion occurs.
  Under such an error model,
  this paper provides a decoding algorithm
  for non-binary quantum codes constructed by Matsumoto and Hagiwara.
\end{abstract}

\section{Introduction}
In quantum coding theory,
{\it insertion} and {\it deletion} \cite{leahy2019quantum}
change the number of \cha{qudits} of a quantum state.
Insertion adds a new \cha{qudit} to a quantum state,
and
deletion loses a \cha{qudit} of a quantum state.

Shibayama \cha{and Ouyang \cite{shibayama2021equivalence}} showed that
any $s_1$ insertions \cha{plus} $s_2$ deletions correcting quantum code
can correct $t_1$ insertions \cha{plus} $t_2$ deletions
\cha{under the modified Knill-Laflamme conditions by Shibayama and Ouyang},
where $s_1 + s_2 = t_1 + t_2$.
For example,
a single-insertion-correcting algorithm \cite{hagiwara2021four} \cha{exists}
for the single-deletion-correcting quantum code \cite{nakayama2020first} and
a multiple-insertion-correcting algorithm \cite{sita2022} \cha{exists}
for the multiple-deletion-correcting quantum codes \cite{matsumoto2022constructions}.
\cha{However,}
there is still no correcting algorithm for a compound error of insertions and deletions.
Even in classical coding theory,
it is \cha{challenging} to construct a correcting algorithm
for a compound error of {\it classical insertions} and {\it deletions}.
Here,
classical insertion adds a new symbol to a sequence\cha{,}
and deletion loses a symbol of a sequence.

Matsumoto and Hagiwara constructed non-binary deletion-correcting quantum codes
\cite{matsumoto2022constructions}.
We provided an insertion-correcting algorithm \cite{sita2022} for the quantum codes.
The purpose of this study is to provide 
an algorithm correcting a compound error of insertions and deletions for the quantum codes.

This paper presents a decoding algorithm correcting {\it single-insertion plus single-deletion}
for the non-binary deletion-correcting quantum codes \cite{matsumoto2022constructions}
constructed by Matsumoto and Hagiwara.
Here,
single-insertion plus single-deletion is a compound error such that
a single insertion occurs and then a single deletion occurs.
To present the decoding algorithm,
we give an algorithm \cha{that}
outputs the candidates of indices of the inserting \cha{qudit}.
By using this,
the received state with single-insertion plus single-deletion
is changed into the state with at most $2$ deletions.
Finally,
we get the message by applying a deletion-correcting algorithm.

\section{Preliminary}
\cha{
  This section introduces
  the notations, error models,
  and
  the non-binary deletion-correcting quantum codes \cite{matsumoto2022constructions}.
}

\subsection{Notations}
Let $\mathbb{Z}$ be the set of all integers.
We denote the cardinality of a set $S$ by $|S|$.
For $a, b, c \in \mathbb{Z}$, 
we define $\setint{a,b} := \{ i \in \mathbb{Z} \mid a \leq i \leq b \}$,
$\setint{b} := \setint{1,b}$, and
$\binom{\setint{b}}{c} := \{ S \subseteq \setint{b} \mid |S|=c \}$.

We denote by $\mathbb{Z}_q^n$ the set of all sequences of length $n$
over a $q$-ary alphabet $\mathbb{Z}_q := \{ 0, 1, ..., q-1 \}$.
We call $\bm{y} \in \mathbb{Z}_q^m$ a {\it subsequence} of $\bm{x} \in \mathbb{Z}_q^n$
if there exists $1 \leq s_1 < s_2 < \cdots < s_m \leq n$
such that $\bm{y} = ( x_{s_1}, x_{s_2}, ..., x_{s_m} )$ for some $m \leq n$.
Then,
we denote the subsequence $( x_{s_1}, x_{s_2}, ..., x_{s_m} )$ of $\bm{x}$
by $\bm{x}_{\{ s_1, s_2, ..., s_m \}}$.
In addition,
$\bm{x}_{\emptyset}$ stands for the empty word.

For $x \in \mathbb{Z}_l$,
let $\ket{x}$ be a column complex vector of dimension $l$
whose only the \cha{$(x+1)$}th component is $1$ and the other components are $0$.
For $\bm{x} = (x_1, x_2, ..., x_n) \in \mathbb{Z}_l^n$,
define $\ket{\bm{x}} := \ket{x_1} \otimes \ket{x_2} \otimes \cdots \otimes \ket{x_n}$
where $\otimes$ is the tensor product.
Let $\text{tr}(\mat{A})$ be the trace of a square matrix $\mat{A}$,
and $\mat{B}^{\dagger}$ be the adjoint matrix of a matrix $\mat{B}$.
Define $\bra{\bm{x}} := \ket{\bm{x}}^{\dagger}$.

We denote the set of all density matrices of $n$ $l$-adic \cha{qudits} by $\State{l}{n}$.
We call a quantum state $\rho \in \State{l}{n}$ {\it pure}
if there exists a vector $\ket{\mu}$ such that $\rho = \ket{\mu}\bra{\mu}$.
In such a case,
we write $\ket{\mu}$ for the pure state $\rho$.
Denote the set of all pure states in $\State{l}{n}$ by $\Pure{l}{n}$.

A {\it \cha{projective measurement}} is defined
by a set $\{\mat{M}_k\}$ of projection matrices $\mat{M}_k$
such that \cha{$\sum_k \cha{\mat{M}_k}$ equals the identity matrix}.
Upon measuring a quantum state $\rho$ by $\{\mat{M}_k\}$,
the probability of getting result $k$ is $\text{tr}(\cha{\mat{M}_k}\rho)$,
and the state after the quantum measurement is $\mat{M}_k\rho\mat{M}_k^{\dagger} / \text{tr}(\cha{\mat{M}_k}\rho)$ \cite{Qcomp_Qinf}.

\subsection{Classical Deletion and Insertion}
Classical deletion changes a symbol to the empty word\cite{levenshtein1966binary}.
For a sequence $\bm{x} \in \mathbb{Z}_q^n$ and
a set of {\it deletion indices} $S \in \binom{\setint{n}}{t}$,
the $t$-{\it classical deletions} $D_S(\bm{x})$ are defined as \cha{follows:}
\begin{align*}
  D_S( \bm{x} )
  &=
  \bm{x}_{\setint{n} \setminus S}.
\end{align*}
For example,
$D_{\{2,4\}}((0,1,2,0)) = (0,2)$.

Classical insertion changes the empty word to a symbol \cite{levenshtein1966binary}.
For a sequence $\bm{x}=(x_i) \in \mathbb{Z}_q^n$, an index $s \in \setint{n+1}$,
and a symbol $\lambda \in \mathbb{Z}_q$,
we define $I_{s, \lambda} : \mathbb{Z}_q^n \to \mathbb{Z}_q^{(n+1)}$ as
\begin{align*}
  I_{s, \lambda}(\bm{x})
  &=
  (x_1, x_2, ..., x_{s-1}, \lambda, x_{s}, ..., x_n).
\end{align*}
Then,
for a sequence $\bm{x} \in \mathbb{Z}_q^n$,
a list of {\it insertion symbols} $\Lambda=(\lambda_1, \lambda_2, ..., \lambda_t)\in \mathbb{Z}_q^t$,
and
a set of {\it insertion indices} $S=\{s_1, s_2, ..., s_t\} \in \binom{\setint{n+t}}{t}$
with $s_1 < s_2 < \cdots < s_t$,
the $t$-{\it classical insertions} $I_{S,\Lambda}(\bm{x})$ are defined as
\begin{align*}
  I_{S,\Lambda}(\bm{x})
  &:=
  I_{s_t,\lambda_t} \circ \cdots \circ I_{s_2,\lambda_2} \circ I_{s_1,\lambda_1}(\bm{x}).
\end{align*}
For example,
$I_{\{1,6\},(\textcolor{red}{2},\textcolor{red}{0})}((0,1,2,0))= (\textcolor{red}{2},0,1,2,0,\textcolor{red}{0})$.

Define
\begin{align*}
  \mathbb{D}^{(t)}(\bm{x})
  &:=
  \left\{
  D_S(\bm{x})
  \middle|
  S \in \tbinom{\setint{n}}{t}
  \right\},\\
  \mathbb{L}^{(t)}(\bm{x})
  &:=
  \left\{
  D_S \circ I_{T,\Lambda}(\bm{x})
  \middle|
  S,T \in \tbinom{\setint{n+t}}{t}, \Lambda \in \mathbb{Z}_q^t
  \right\}
  \setminus \{\bm{x}\}.
\end{align*}
In words,
$\mathbb{D}^{(t)}(\bm{x})$ is
the set of sequences after $t$ classical deletions to $\bm{x} \in \mathbb{Z}_q^n$,
and
$\mathbb{L}^{(t)}(\bm{x})$ is
the set of sequences after $t$ classical insertions and $t$ classical deletions
to $\bm{x} \in \mathbb{Z}_q^n$ except $\bm{x}$.
  
\subsection{Indel Distance \cite{levenshtein1966binary}}\label{subsec:insdel_saiki}
For $\bm{x} \in \mathbb{Z}_q^n$ and $ \bm{y} \in \mathbb{Z}_q^m$,
the {\it indel distance} $d( \bm{x}, \bm{y} )$ of $\bm{x}$ and $\bm{y}$ is defined by
the smallest number of classical deletions and insertions
that transform $\bm{x}$ into $\bm{y}$ \cite{levenshtein1966binary}.
Since the indel distance is one of the edit distance \cite{wagner1974string},
it is calculated by the algorithm in \cite{wagner1974string},
which is shown in Algorithm \ref{proc:indel}.
The output $\mat{H}_{\bm{x}, \bm{y}} = (h_{i,j})$
satisfies $h_{i,j} = d(\bm{x}_{\setint{i}}, \bm{y}_{\setint{j}})$
for any $i \in \setint{0,n}$ and $j \in \setint{0,m}$ \cite{wagner1974string}.

\begin{algo}\label{proc:indel}~\\
  {\bf Input: }Sequences $\bm{x} \in \mathbb{Z}_q^n$, $ \bm{y} \in \mathbb{Z}_q^m$\\
  {\bf Output: }$(n+1) \times (m+1)$ matrix $\mat{H}_{\bm{x}, \bm{y}} = (h_{i,j})$
  \begin{enumerate}
  \item Set $h_{i,0} \gets i$ and $h_{0,j} \gets j$
    for all $i \in \setint{0,n}$ and $j \in \setint{0,m}$.
  \item Calculate
    \begin{align}
      h_{i,j}
      \gets
      \min
      \left\{
      \begin{aligned}
        &h_{i,j-1}+1,\\
        &h_{i-1,j}+1,\\
        &h_{i-1,j-1} + 2 \1 \{ x_i \neq y_j \}
      \end{aligned}
      \right\}
      \label{eq:matrix}
    \end{align}
    from $h_{1,1}$ to $h_{n,m}$ in order,
    where $\1\{ x_i \neq y_j \}$ is $1$ if $x_i \neq y_j$ and $0$ otherwise.
  \end{enumerate}
\end{algo}

\subsection{Set of Sequences Detecting Deletion Indices}
\label{ssec:p_dash}
Let $\mathbb{E}_{n,q,t}$ be a set of sequences $\bm{x} \in \mathbb{Z}_q^n$
that can uniquely detect $t$ deletion indices $S \in \binom{\setint{n}}{t}$ from $D_S(\bm{x})$.
For example,
$\mathbb{E}_{n,q,t}$ includes
the sequence \cite{matsumoto2022constructions} repeating the symbols from $0$ to $t$ such as
$(0, 1 ,..., t,0,1, ..., t,0, ...) =: \M \in \mathbb{Z}_{t+1}^n$.
\cha{
  We call the sequence $\M$
  {\it the monotonically increasing periodic sequence on $\mathbb{Z}_{t+1}^n$.}
}
More precisely,
$\M = (m_i)$ satisfies $m_i = (i-1) \% (t+1)$ for any $i \in \setint{n}$,
where $(i-1) \% (t+1)$ is the remainder when $(i-1)$ is divided by $(t+1)$.

\subsection{Unitary error and Deletion/Insertion}
\cha{
  A {\it single-unitary error} changes
  from a quantum state $\rho = \sum_{ \bm{x},\bm{y} \in \mathbb{Z}_l^n } c_{ \bm{x},\bm{y} } \ket{\bm{x}}\bra{\bm{y}} \in \State{l}{n}$ to
  \begin{align*}
    \sum_{ \bm{x},\bm{y} \in \mathbb{Z}_l^n }
    c_{ \bm{x},\bm{y} }
    \ket{\bm{x}_{\setint{i-1}}}\bra{\bm{y}_{\setint{i-1}}} \otimes&\\
    \mat{U}\ket{x_i}\bra{y_i}\mat{U}^\dagger \otimes&
    \ket{\bm{x}_{\setint{i+1,n}}}\bra{\bm{y}_{\setint{i+1,n}}},
  \end{align*}
  for some $l \times l$ unitary matrix $\mat{U}$ and some index $i \in \setint{n}$.
}

For a quantum state $\rho = \sum_{ \bm{x},\bm{y} \in \mathbb{Z}_l^n } c_{ \bm{x},\bm{y} } \ket{\bm{x}}\bra{\bm{y}} \in \State{l}{n}$ and an index $j \in \setint{n}$,
the {\it partial trace} $\text{Tr}_j : \State{l}{n} \to \State{l}{(n-1)}$ is defined as
\begin{align*}
  \text{Tr}_j( \rho )
  =
  \sum_{ \bm{x},\bm{y} \in \mathbb{Z}_l^n } c_{ \bm{x},\bm{y} }
  \text{tr}( \ket{x_j}\bra{y_j} )
  \ket{D_{\{j\}}(\bm{x})}\bra{D_{\{j\}}(\bm{y})}.
\end{align*}
Then,
for a quantum state $\rho \in \State{l}{n}$ and
a set of {\it deletion indices} $J = \{ j_1, j_2, ..., j_t \} \in \binom{\setint{n}}{t}$
with $j_1 < j_2 < \cdots < j_t$,
the $t$-{\it deletions} $\mathcal{D}_J(\rho)$ are defined as 
\begin{align*}
  \mathcal{D}_J(\rho)
  :=
  \text{Tr}_{j_1}\circ \text{Tr}_{j_2} \circ\cdots\circ \text{Tr}_{j_t}(\rho).
\end{align*}
We refer to the error model
where the receiver gets the received state with deletions and knows the deletion indices
as {\it \cha{erasures}}.

\cha{
  For $\rho \in \State{l}{n}$,
  $t$-insertions at a set of {\it insertion indices} $J \in \binom{\setint{n+t}}{t}$ is a change
  from $\rho$
  to $\rho' \in \{ \sigma \in \State{l}{(n+t)} \mid \mathcal{D}_J(\sigma)=\rho \} =: \mathcal{B}_J(\rho)$.
  If $\rho$ is pure,
  $\rho' \in \mathcal{B}_P(\rho)$ is denoted by the following.
  \begin{theo}[Theorem $3.3$ of \cite{sibayama_single_ins_def}]
    \label{theo:after_ins}
    Consider
    a quantum state $\rho = \sum_{\bm{x},\bm{y} \in \mathbb{Z}_l^n} c_{\bm{x},\bm{y}}\ket{\bm{x}}\bra{\bm{y}} \in \State{l}{n}$ and
    a permutation $\tau$ on $\setint{1,n}$.
    By abuse of notation,
    we define the {\it index permutation} $\tau(\rho) \in \State{l}{n}$ for $\rho$ by
    \begin{align*}
      \tau(\rho)
      :=
      \sum_{\bm{x},\bm{y} \in \mathbb{Z}_l^n}
      c_{\bm{x},\bm{y}}
      \ket{x_{\tau(1)} \cdots x_{\tau(n)}}\bra{y_{\tau(1)} \cdots y_{\tau(n)}}.
    \end{align*}
    In addition,
    for $t,n \in \mathbb{Z}^+$ and
    $J = \{j_1,j_2,...,j_t\} \subset \setint{1,n+t}$ with $j_1 < j_2 < \cdots < j_t$,
    let $\tau_J$ be the permutation on $\setint{1,n+t}$
    such that $\tau_J(i)=j_i$ for $i \in \setint{1,t}$ and
    $i_1 < i_2 \Rightarrow \tau_J(i_1) < \tau_J(i_2)$ for $i_1,i_2 \in \setint{t+1,n+t}$.
    Then,
    if $\rho \in \State{l}{n}$ is pure,
    any quantum state $\rho' \in \mathcal{B}_J(\rho)$ is denoted
    by $\rho' = \tau_J (\sigma \otimes \rho)$
    for some $\sigma \in \State{l}{t}$.
  \end{theo}  
}

\cha{For $\rho,\sigma$, and $J$,
  we denote $\tau_J (\sigma \otimes \rho) \in \mathcal{B}_J(\rho)$
  by $\mathcal{I}_{J,\sigma}(\rho)$.
}
Let $\mathscr{L}^{(t)}(\cha{\rho})$ be
the set of quantum states after $t$-insertion plus $t$-deletion to $\cha{\rho} \in \Pure{l}{n}$,
i.e.,
\begin{align*}
  \mathscr{L}^{(t)}(\cha{\rho})
  &=
  \left\{
  \mathcal{D}_{J_1} \circ \mathcal{I}_{J_2,\pi}(\cha{\rho})
  \middle|
  J_1, J_2 \in \tbinom{\setint{n+t}}{t}, \pi \in \State{l}{t}
  \right\}.
\end{align*}

\subsection{Non-binary Deletion-Correcting Quantum Codes \cite{matsumoto2022constructions}}
\label{subsec:matu+hagi_code}
Let $\mathcal{C}_{n,l,t} \subseteq \Pure{l(t+1)}{n}$
be the non-binary $t$-deletion-correcting quantum code \cite{matsumoto2022constructions}
constructed by Matsumoto and Hagiwara.
The code $\mathcal{C}_{n,l,t}$ is obtained
by a $t$-erasure-correcting quantum code $\mathcal{C}_0 \subseteq \Pure{l}{n}$.
Define the map $\eta_i$ as $\eta_i:\ket{j} \mapsto \ket{j(t+1)+i}$ for $j \in \setint{0,l-1}$.
Then,
a codeword of $\mathcal{C}_{n,l,t}$ is given
by applying $\eta_{(i-1) \% (t+1)}$ to the $i$th \cha{qudit} of a codeword of $\mathcal{C}_0$
for all $i \in \setint{n}$.

The \cha{projective measurement} $\mathcal{M}_t$ for $\mathcal{C}_{n,l,t}$ is defined by
\begin{align*}
  \left\{
  \mat{M}_k := \sum_{j=0}^{l-1}\ket{ j(t+1)+k }\bra{ j(t+1)+k }
  \middle|
  k \in \setint{0,t}
  \right\}.
\end{align*}
Upon measuring the $i$th \cha{qudit} of a codeword of $\mathcal{C}_{n,l,t}$ by $\mathcal{M}_t$,
we get the result $(i-1) \% (t+1)$ with the probability $1$,
and the state does not change \cite{matsumoto2022constructions}.
This implies that
\cha{
  the sequence of the results coincides with
  the monotonically increasing periodic sequence $\M$ on $\mathbb{Z}_{t+1}^n$.
}

\section{Decoding Algorithm Correcting Single-Insertion Plus Single-Deletion}
\subsection{Decoding Algorithm}\label{subsec:decoder}
Suppose $t \ge 2$.
Algorithm \ref{proc:decoder} \cha{corrects} single-insertion plus single-deletion for $\mathcal{C}_{n,l,t}$.

\begin{figure}[t]
  \begin{algorithm}[H]
    \caption{~}
    \label{algo:fi(p1)} 
    \begin{algorithmic}[1]    
      \REQUIRE Sequences $\bm{x} \in \mathbb{Z}_{q}^n$, $\bm{y} \in \mathbb{Z}_q^m$,
      the output $\mat{H}_{\bm{x},\bm{y}}$ of Algorithm \ref{proc:indel}
      \ENSURE A subset $S_1 \subseteq \setint{m}$
      \STATE $i \gets n$, $j \gets m$, $S_1 \gets \emptyset$
      \WHILE{\cha{$i \ge 1$ \OR $j \ge 1$}}
      \IF{$i \ge 1$ and $h_{i,j} = h_{i-1,j}+1$}
      \STATE $i \gets i-1$
      \ELSIF{$i,j \ge 1$ and $x_i = y_j$}
      \STATE $i \gets i-1, j \gets j-1$
      \ELSIF{$j \ge 1$ and $h_{i,j}=h_{i,j-1}+1$}
      \STATE $S_1 \gets S_1 \cup \{ j \}$, $j \gets j-1$
      \ENDIF
      \ENDWHILE
    \end{algorithmic}
  \end{algorithm}
\end{figure}

\begin{figure}[t]
  \begin{algorithm}[H]
    \caption{~}
    \label{algo:fi(p2)} 
    \begin{algorithmic}[1]    
      \REQUIRE Sequences $\bm{x} \in \mathbb{Z}_{q}^n$, $\bm{y} \in \mathbb{Z}_q^m$,
      the output $\mat{H}_{\bm{x},\bm{y}}$ of Algorithm \ref{proc:indel}
      \ENSURE A subset $S_2 \subseteq \setint{m}$
      \STATE $i \gets n$, $j \gets m$, $S_2 \gets \emptyset$
      \WHILE{\cha{$i \ge 1$ \OR $j \ge 1$}}
      \IF{$j \ge 1$ and $h_{i,j}=h_{i,j-1}+1$}
      \STATE $S_2 \gets S_2 \cup \{ j \}$, $j \gets j-1$
      \ELSIF{$i,j \ge 1$ and $x_i = y_j$}
      \STATE $i \gets i-1, j \gets j-1$
      \ELSIF{$i \ge 1$ and $h_{i,j} = h_{i-1,j}+1$}
      \STATE $i \gets i-1$
      \ENDIF
      \ENDWHILE
    \end{algorithmic}
  \end{algorithm}
\end{figure}

\begin{algo}\label{proc:decoder} ~ \\
  {\bf Input: }A received state $\rho \in \mathscr{L}^{(1)}(\ket{\psi})$\\
  {\bf Output: }A quantum state $\hat{\pi} \in \Pure{l}{u}$
  \begin{enumerate}
  \item Perform $\mathcal{M}_t$ on each \cha{qudit} in $\rho$.
    Let $\sigma$ be the quantum state after the measurement
    and $\bm{r}$ be the sequence of the results.
    \cha{If $\bm{r} = \M$, go to step $2$.
      Otherwise, go to step $3$.}
  \item Output a quantum state 
    obtained by applying a single-unitary-error-correcting algorithm of $\mathcal{C}_0$ to $\sigma$.
  \item Get $\mat{H}_{\M,\bm{r}}$
    by Algorithm \ref{proc:indel} for input $\M$ and $\bm{r}$.
  \item Make $S_1$ (resp. $S_2$) by Subroutine \ref{algo:fi(p1)} (resp. \ref{algo:fi(p2)})
    for input $\M$, $\bm{r}$, and $\mat{H}_{\M,\bm{r}}$.
  \item Get $\mathcal{D}_{ S_1 \cup S_2 }( \sigma )$.
  \item Output a quantum state
    obtained by applying a deletion-correcting algorithm of $\mathcal{C}_{n,l,t}$
    to $\mathcal{D}_{ S_1 \cup S_2 }( \sigma )$.
  \end{enumerate}
\end{algo}

\begin{rem}
  Subroutine \ref{algo:fi(p2)} is obtained by swapping the lines $3,4$ and the lines $7,8$
  of Subroutine \ref{algo:fi(p1)}.
\end{rem}

\begin{rem}
  Since any $2t$-erasure-correcting quantum code corrects
  $t$ unitary errors \cite{gottesman2005quantum},
  there exists a single-unitary-error-correcting algorithm of $\mathcal{C}_0$.
\end{rem}

\subsection{Justification}\label{subsec:deco_jus}
This section proves that
Algorithm \ref{proc:decoder} corrects
any single-insertion plus single-deletion for $\mathcal{C}_{n,l,t}$.
The main statement is the following.

\begin{theo}\label{theo:pi}
  Let $\ket{\mu} \in \Pure{l}{u}$ be a message and
  $\ket{\psi} \in \mathcal{C}_{n,l,t}$ be the corresponding codeword.
  Then,
  if $t \ge2$,
  the output $\hat{\pi}$ of Algorithm \ref{proc:decoder}
  for input $\rho \in \mathscr{L}^{(1)}(\ket{\psi})$ satisfies
  \begin{align*}
    \hat{\pi}
    &=
    \ket{\mu}.
  \end{align*}
\end{theo}

To prove this, we use Theorems \ref{theo:step1}, \ref{theo:r=m_result_class},
and Corollaries \ref{theo:r_neq_m_result}, \ref{coro:r=m_result}.

\begin{theo}\label{theo:step1}
  Assume the received state $\rho$ is in $\mathscr{L}^{(1)}(\ket{\psi})$
  for $\ket{\psi} \in \mathcal{C}_{n,l,t}$.
  Let $\sigma$ and $\bm{r}$ be the quantum state and the sequence
  obtained by step $1$ of Algorithm \ref{proc:decoder}.
  Then,
  $\sigma \in \mathscr{L}^{(1)}(\ket{\psi})$ and $\bm{r} \in \mathbb{L}^{(1)}(\M)$ hold
  and
  the insertion (resp. deletion) index of $\rho$ coincides with
  the insertion (resp. deletion) index of $\sigma$ and $\bm{r}$.
\end{theo}

\begin{IEEEproof}
  Upon measuring the $i$th \cha{qudit} of $\ket{\psi} \in \mathcal{C}_{n,l,t}$ by $\mathcal{M}_t$,
  we get the result $(i-1) \% (t+1)$ with the probability $1$
  and the state does not change.
  On the other hand,
  upon measuring the insertion \cha{qudit} by $\mathcal{M}_t$,
  we get some result $k \in \mathbb{Z}_{t+1}$
  with the probability $\text{tr}( \cha{\mat{M}_k} \rho_0 )$
  and the state changes to $\mat{M}_k \rho_0 \mat{M}_k^{\dagger} / \text{tr}( \cha{\mat{M}_k} \rho_0)$.
  Therefore,
  $\sigma \in \mathscr{L}^{(1)}(\ket{\psi})$ and $\bm{r} \in \mathbb{L}^{(1)}(\M)$ hold
  and
  the insertion (resp. deletion) index of $\rho$ coincides with
  the insertion (resp. deletion) index of $\sigma$ and $\bm{r}$.
\end{IEEEproof}

\begin{coro}\label{theo:r_neq_m_result}
  For $\ket{\psi} \in \mathcal{C}_{n,l,t}$ and $\rho \in \mathscr{L}^{(1)}(\ket{\psi})$,
  define $\sigma$ and $\bm{r}$ as in Theorem \ref{theo:step1}.
  Then,
  if $\bm{r} = \M$,
  $\sigma$ is a quantum state after a single unitary error to $\ket{\psi}$.
\end{coro}

Theorem \ref{theo:r=m_result_class} means that
the insertion index of $\bm{r}$ is either
the output of Subroutine \ref{algo:fi(p1)} or \ref{algo:fi(p2)}.
Theorem \ref{theo:r=m_result_class} will be proven in Section \ref{sec:theo}.

\begin{theo}\label{theo:r=m_result_class}
  For $\bm{x} \in \mathbb{E}_{n,q,t}$ and $\bm{y} \in \mathbb{L}^{(1)}(\bm{x})$,
  define $J := \{ j \in \setint{n} \mid D_{\{j\}}(\bm{y}) \in \mbb{D}^{(1)}(\bm{x}) \}$.
  Let $S_1$ (resp. $S_2$) be the output of
  Subroutine \ref{algo:fi(p1)} (resp. \ref{algo:fi(p2)})
  for input $\bm{x}$, $\bm{y}$, and $\mat{H}_{\bm{x}, \bm{y}}$.
  Then,
  if $t \ge 2$,
  $S_1$ and $S_2$ satisfy
  \begin{align*}
    &J = S_1 \cup S_2,
    &&|S_1 \cup S_2| \leq 2.
  \end{align*}
\end{theo}

Corollary \ref{coro:r=m_result} immediately follows
from Theorems \ref{theo:step1}, \ref{theo:r=m_result_class}.

\begin{coro}\label{coro:r=m_result}
  Assume the received state $\rho$ is in $\mathscr{L}^{(1)}(\ket{\psi})$
  for $\ket{\psi} \in \mathcal{C}_{n,l,t}$.
  Define $\sigma$ and $\bm{r}$ as in Theorem \ref{theo:step1}.
  Similarly,
  define $S_1$ and $S_2$ by step $4$ of Algorithm \ref{proc:decoder}.
  Then,
  if $\bm{r} \neq \M$,
  $\mathcal{D}_{S_1 \cup S_2 }(\sigma)$ is a quantum state after at most $2$ deletions to $\ket{\psi}$.
\end{coro}

From the above,
we prove Theorem \ref{theo:pi}.

\begin{IEEEproof}[Proof of Theorem \ref{theo:pi}]
  If $\bm{r} = \M$,
  from Corollary \ref{theo:r_neq_m_result},
  the quantum state outputed by step $2$ coincides with the message of $\ket{\psi}$.
  If $\bm{r} \neq \M$,
  from Corollary \ref{coro:r=m_result},
  the quantum state obtained by step $6$ coincides with the message of $\ket{\psi}$.
\end{IEEEproof}

\section{Proof of Theorem \ref{theo:r=m_result_class}}\label{sec:theo}
Section \ref{subsubsec:proof} gives the lemmas for the proof of Theorem \ref{theo:r=m_result_class}.
Section \ref{subsec:lem3_proof} proves Theorem \ref{theo:r=m_result_class}.

\subsection{Lemmas to prove Theorem \ref{theo:r=m_result_class}}
\label{subsubsec:proof}
For a given digraph $\mtt{G} = (\mtt{V}(\mathtt{G}), \mtt{A}(\mtt{G}))$,
a path $\mtt{P}$ is represented by a list of vertices $\mtt{v}_{i_1} \mtt{v}_{i_2} ... \mtt{v}_{i_k}$
where $(\mtt{v}_{i_j}, \mtt{v}_{i_{j+1}}) \in \mtt{A}(\mtt{G})$ for $j = 1, 2, ..., k-1$.
Suppose $\mtt{P} = \mtt{v}_{i_1} \mtt{v}_{i_2} ... \mtt{v}_{i_k}$.
Then,
we call that \cha{the} vertex $\mtt{v}_{i_j}$ is included in $\mtt{P}$.
Similarly,
\cha{the} arc $(\mtt{v}_{i_j}, \mtt{v}_{i_{j+1}})$ is included in $\mtt{P}$.
By abuse of notation,
$\mtt{v} \in \mtt{P}$ (resp. $\mtt{a} \in \mtt{P}$) represents that
vertex $\mtt{v}$ (resp. arc $\mtt{a}$) is included in $\mtt{P}$.

\begin{defi}
  Fix $\bm{x} \in \mbb{Z}_q^n$ and $\bm{y} \in \mbb{Z}_q^m$.
  Define $\mat{H}_{\bm{x}, \bm{y}} = (h_{i,j})$ as in Algorithm \ref{proc:indel}.
  Then,
  define the digraph $\mtt{G}_{\bm{x}, \bm{y}} =(\digrV, \digrA)$ as
  \begin{align*}
    \digrV
    &:= 
    \{ \mtt{v}_{i,j} \mid i \in \setint{0,n}, j \in \setint{0,m} \}\\
    \digrA
    &:=
    \mtt{A}_1 \cup \mtt{A}_2 \cup \mtt{A}_3,
  \end{align*}
  where
  \begin{align*}
    \mtt{A}_1
    &:=
    \{
    (\mtt{v}_{i,j}, \mtt{v}_{i-1,j}) 
    \mid
    i \in \setint{n},
    j \in \setint{0,m},
    h_{i,j} = h_{i-1,j} + 1
    \},\\
    \mtt{A}_2
    &:=
    \{
    (\mtt{v}_{i,j},\mtt{v}_{i-1,j-1})
    \mid
    i \in \setint{n},
    j \in \setint{m},
    x_{i} = y_{j}
    \},\\
    \mtt{A}_3
    &:=
    \{
    (\mtt{v}_{i,j}, \mtt{v}_{i,j-1}) 
    \mid
    i \in \setint{0,n},
    j \in \setint{m},
    h_{i,j} = h_{i,j-1} + 1
    \}.
  \end{align*}
  Moreover,
  for $i=1,2,3$,
  we refer to arcs in $\mtt{A}_i$ as {\it type} $i$.
\end{defi}

A vertex $\mtt{v}_{i,j} \in \digrV$ corresponds
to $h_{i,j}$ of $\mat{H}_{\bm{x}, \bm{y}} = (h_{i,j})$.
In addition,
an arc of type $1$, $2$, and $3$ corresponds to
an arrow pointing up, left up, and left in Fig. \ref{fig:example_digraph},
respectively.

The following lemma means that
a path from $\mtt{v}_{n,m}$ to $\mtt{v}_{0,0}$ exists in $\mtt{G}_{\bm{x}, \bm{y}}$.

\begin{lem}\label{lem:exist_path}
  For any $\bm{x} \in \mbb{Z}_q^n$, $\bm{y} \in \mbb{Z}_q^m$, and
  any $\mtt{v}_{i,j} \in \digrV \setminus \{\mtt{v}_{0,0}\}$,
  $(\mtt{v}_{i,j}, \mtt{w}) \in \digrA$ holds with some $\mtt{w} \in \digrV$.
\end{lem}
\begin{IEEEproof}
  If $x_{i} = y_{j}$,
  $(\mtt{v}_{i,j},\mtt{v}_{i-1,j-1}) \in \mtt{A}_2$ holds.
  If $x_{i} \neq y_{j}$,
  from the triangle inequality,
  we get
  \begin{align*}
    h_{i,j-1} +1
    &=
    d(\bm{x}_{\setint{i}}, \bm{y}_{\setint{j-1}}) +1\\
    &\leq
    d(\bm{x}_{\setint{i}}, \bm{x}_{\setint{i-1}}) + d(\bm{x}_{\setint{i-1}}, \bm{y}_{\setint{j-1}}) +1\\
    &=
    d(\bm{x}_{\setint{i-1}}, \bm{y}_{\setint{j-1}}) +2\\
    &=
    h_{i-1,j-1} + 2 \1 \{ x_{i} \neq y_{j} \}.
  \end{align*}
  Hence,
  from \eqref{eq:matrix},
  we have $h_{i,j} = \min\{ h_{i-1,j}+1, h_{i,j-1}+1 \}$.
  This yields 
  $(\mtt{v}_{i,j},\mtt{v}_{i-1,j}) \in \mtt{A}_1$ or $(\mtt{v}_{i,j},\mtt{v}_{i,j-1}) \in \mtt{A}_3$.
\end{IEEEproof}

For $\bm{x} \in \mbb{Z}_q^n$ and $\bm{y} \in \mbb{Z}_q^m$,
let $\Pset$ be
the set of paths in $\mtt{G}_{\bm{x}, \bm{y}}$ from $\mtt{v}_{n,m}$ to $\mtt{v}_{0,0}$.

\begin{defi}\label{def:p_top}  
  For $\bm{x} \in \mbb{Z}_q^n$ and $\bm{y} \in \mbb{Z}_q^m$,
  $\mtt{P} \leq \mtt{Q}$ represents 
  the relation between $\mtt{P} \in \Pset$ and $\mtt{Q} \in \Pset$ such that
  \begin{align*}
    \min\{ i \in \setint{0,n} \mid \mtt{v}_{i,j} \in \mtt{P}\} 
    &\leq 
    \min\{ i \in \setint{0,n} \mid \mtt{v}_{i,j} \in \mtt{Q}\},\\
    \max\{ i \in \setint{0,n} \mid \mtt{v}_{i,j} \in \mtt{P}\} 
    &\leq 
    \max\{ i \in \setint{0,n} \mid \mtt{v}_{i,j} \in \mtt{Q}\},
  \end{align*}
  for any $j \in \setint{0,m}$.
  Then,
  we denote the maximum (resp. minimum) element of the poset $(\Pset, \leq)$
  by $\Ptop$ (resp. $\Pbot$).
\end{defi}

\begin{rem}\label{rem:ptop}
  In Subroutine \ref{algo:fi(p1)},
  initially $(i,j) = (n,m)$ is set,
  $(i,j)$ is updated in each iteration,
  and finally $(i,j)$ becomes $(0,0)$.
  Let $k_t$ represent the pair $(i,j)$ at the end of the $t$th iteation
  and $\tau$ be the total number of iteations.
  Here,
  $k_0 = (n,m)$ and $k_{\tau}=(0,0)$.
  Then,
  $\mtt{v}_{k_0} \mtt{v}_{k_1} \mtt{v}_{k_2} \cdots \mtt{v}_{k_\tau}$ forms
  a path in $\mtt{G}_{\bm{x}, \bm{y}}$ and is in $\Pset$.
  Moreover,
  this path equals $\Pbot$.
  Similarly,
  the path generated by Subroutine \ref{algo:fi(p2)} equals $\Ptop$.
\end{rem}
  
\begin{exam}\label{exam:leq}
  Assume $\bm{x}=(0,1,2)$ and $\bm{y}=(1,1,2)$.
  Fig. \ref{fig:example_digraph} depicts $\mathtt{G}_{\bm{x}, \bm{y}}$
  where $\mtt{v}_{i,j} \in \digrV$ is represented by the value of $h_{i,j}$.  
  Then,
  $\Ptop$ and $\Pbot$ are given as follows.
  \begin{align*}
    &\Pbot
    =
    \mtt{v}_{3,3} \mtt{v}_{2,2} \mtt{v}_{1,1} \mtt{v}_{0,1} \mtt{v}_{0,0},
    &&\Ptop
    =
    \mtt{v}_{3,3} \mtt{v}_{2,2} \mtt{v}_{2,1} \mtt{v}_{1,0} \mtt{v}_{0,0}.
  \end{align*}
  The path $\Ptop$ (resp. $\Pbot$) is the maximum (resp. minimum) element of $(\Pset, \leq)$
  shown as in Fig. \ref{fig:example_ptop}.

  \begin{figure}[tt]
    \begin{minipage}[b]{0.45\linewidth}
      \centering
      \begin{tabular}{c|c|c*{6}{@{}c}|}
        &$j$& 0   &&1   &&2   &&3\\\hline
        $i$&$\bm{x} \backslash \bm{y}$& &&1   &&1  &&2\\
        \hline
        \vspace{-1mm}
        0& &0 &$\gets$&1 &$\gets$&2 &$\gets$&3\\
        \vspace{-1mm}
        &&$\uparrow$ &&$\uparrow$ &&$\uparrow$ &&$\uparrow$\\
        \vspace{-1mm}
        1&0 &1 &$\gets$&2 &$\gets$&3 &$\gets$&4\\
        \vspace{-1mm}
        &&$\uparrow$ &$\nwarrow$& &$\nwarrow$& &&\\
        \vspace{-1mm}
        2&1 &2 &&1 &$\gets$&2 &$\gets$&3\\
        \vspace{-1mm}
        &&$\uparrow$ &&$\uparrow$ &&$\uparrow$ &$\nwarrow$&\\
        3&2 &3 &&2 &$\gets$&3 &&2\\
        \hline  
      \end{tabular}
      \caption{Digraph $\mtt{G}_{(0,1,2), (1,1,2)}$}
      \label{fig:example_digraph}
    \end{minipage}
    \begin{minipage}[b]{0.45\linewidth}
      \centering
      \begin{tabular}{c|c|c*{6}{@{}c}|}
        &$j$& 0   &&1   &&2   &&3\\\hline
        $i$&$\bm{x} \backslash \bm{y}$& &&1   &&1  &&2\\
        \hline
        \vspace{-1mm}
        0& &0 &$\gets$&1 &&2 &&3\\
        \vspace{-1mm}
        &&$\uparrow$ &&$\uparrow$ &$\Pbot$& &&\\
        \vspace{-1mm}
        1&0 &1 &&2 &&3 &&4\\
        \vspace{-1mm}
        && &$\nwarrow$& &$\nwarrow$& &&\\
        \vspace{-1mm}
        2&1 &2 &&1 &$\gets$&2 &&3\\
        \vspace{-1mm}
        && && &$\Ptop$& &$\nwarrow$&\\
        3&2 &3 &&2 &&3 &&2\\
        \hline  
      \end{tabular}
      \caption{$\Pbot$ and $\Pbot$}
      \label{fig:example_ptop}
    \end{minipage}
  \end{figure}  
\end{exam}

For a given $\mtt{P} \in \Pset$ and $i=1,2,3$,
we denote the set of arcs $\mtt{a} \in \mtt{P}$ of type $i$ by $\mtt{A}_i(\mtt{P})$,
i.e.,
$\mtt{A}_i(\mtt{P}) = \{ \mtt{a} \in \mtt{A}_i \mid \mtt{a} \in \mtt{P}\}$.

\begin{lem}\label{lem:each_size}
  For $\bm{x} \in \mbb{Z}_q^n$ and $\bm{y} \in \mbb{Z}_q^m$,
  any path $\mtt{P} \in \Pset$ satisfies
  \begin{align}
    |\mtt{A}_1(\mtt{P})|
    &=
    (d(\bm{x}, \bm{y}) +n -m)/2,\label{eq:alpha_size}\\
    |\mtt{A}_2(\mtt{P})|
    &=
    (-d(\bm{x}, \bm{y}) +n +m)/2,\label{eq:beta_size}\\
    |\mathtt{A}_3(\mtt{P})|
    &=
    (d(\bm{x}, \bm{y}) -n +m)/2.\label{eq:gamma_size}
  \end{align}  
\end{lem}

\begin{IEEEproof}
  The triangle inequality leads
  \begin{align*}
    h_{i,j}
    &\leq
    d(\bm{x}_{\setint{i-1}}, \bm{x}_{\setint{i}}) + d(\bm{x}_{\setint{i}}, \bm{y}_{\setint{j-1}})
    =
    1 + h_{i,j-1}.
  \end{align*}
  Similarly,
  we get $h_{i,j} \leq 1 + h_{i-1,j}$.
  Hence,
  from \eqref{eq:matrix},
  if $(\mtt{v}_{i,j}, \mtt{v}_{i-1,j-1}) \in \mathtt{A}_2(\mtt{P})$,
  we have
  \begin{align}
    h_{i,j}
    &=
    \min
    \{
    h_{i-1,j}+1,
    h_{i,j-1}+1,
    h_{i-1,j-1}
    \}
    =
    h_{i-1,j-1}\nonumber\\
    \iff 
    0
    &=
    h_{i,j} - h_{i-1,j-1}.
    \label{eq:num_typ_b}
  \end{align}
  Additionally,
  $h_{i,j}$ satisfies
  \begin{align}
    1
    &=
    h_{i,j} - h_{i-1,j}
    &&\text{if } (\mtt{v}_{i,j}, \mtt{v}_{i-1,j}) \in \mathtt{A}_1(\mtt{P}),\label{eq:num_typ_a}\\
    1
    &=
    h_{i,j} - h_{i,j-1}
    &&\text{if } (\mtt{v}_{i,j}, \mtt{v}_{i,j-1}) \in \mathtt{A}_3(\mtt{P}).\label{eq:num_typ_r}
  \end{align}
  Combining \eqref{eq:num_typ_b}, \eqref{eq:num_typ_a}, and \eqref{eq:num_typ_r},
  we get
  \begin{align}
    |\mathtt{A}_1(\mtt{P})| + |\mathtt{A}_3(\mtt{P})|
    &=
    \sum_{(\mtt{v}_{i,j}, \mtt{v}_{i',j'}) \in \mtt{P}}~
    (
    h_{i,j} - h_{i',j'}
    )\nonumber\\
    &=
    h_{n,m} - h_{0,0}
    ~~=
    d(\bm{x}, \bm{y}).
    \label{lem:wa=d}
  \end{align}
  Now,
  it requires $n$ upward moves to go to $\mtt{v}_{0,0}$ from $\mtt{v}_{n,m}$.
  An arc of type $1$ and $2$ gives one upward move.
  Hence,
  the sum of the number of types $1$ and $2$ equals $n$,
  i.e.,
  \begin{align}
    n
    &=
    |\mtt{A}_1(\mtt{P})|+|\mtt{A}_2(\mtt{P})|.\label{eq:n=a+b}
  \end{align}
  Similarly,
  considering leftward moves,
  we have 
  \begin{align}
    m
    &=
    |\mtt{A}_2(\mtt{P})|+|\mtt{A}_3(\mtt{P})|.\label{eq:n=b+r}
  \end{align}
  We obtain \eqref{eq:alpha_size}, \eqref{eq:beta_size}, and \eqref{eq:gamma_size}
  by solving \eqref{lem:wa=d}, \eqref{eq:n=a+b}, and \eqref{eq:n=b+r}.
\end{IEEEproof}

\begin{lem}\label{lem:only_exit}
  For any $\bm{x} \in \mbb{E}_{n,q,t}$,
  if $|j-i| \leq t$ and $x_i = x_j$,
  $i = j$ holds.
\end{lem}

\begin{IEEEproof}
  Assume $i \leq j$ without loss of generality.
  Hypothesize $i \neq j$.
  Then,
  $D_{\setint{i+1,j-1}}(\bm{x}) = (x_1,...,x_i,x_j,...,x_n)$.
  Since $x_i =y_j$,
  we have $D_{\setint{i,j-1}}(\bm{x}) = D_{\setint{i-1,j}}(\bm{x}) \in \mbb{D}^{(j-i)}(\bm{x})$.
  This implies that $\bm{x}$ cannot detect $j-i \leq t$ deletion indices.
  This contradicts $\bm{x} \in \mbb{E}_{n,q,t}$.
\end{IEEEproof}

For $\bm{x} \in \mbb{Z}_q^n$, $\bm{y} \in \mbb{Z}_q^m$ and $\mtt{P} \in \Pset$,
define 
\begin{align*}
  f_D(\mtt{P})
  &:=
  \{
  i \in \setint{n}
  \mid
  \exists j \in \setint{0,m} ~(\mtt{v}_{i,j},\mtt{v}_{i-1,j}) \in \mtt{A}_1(\mtt{P})
  \},\\
  f_I(\mtt{P})
  &:=
  \{
  j \in \setint{m}
  \mid
  \exists i \in \setint{0,n} ~(\mtt{v}_{i,j},\mtt{v}_{i,j-1}) \in \mtt{A}_3(\mtt{P})
  \}.
\end{align*}
\begin{rem}
  Note that any $\mtt{P} \in \Pset$ satisfies
  \begin{align}
    &|f_D(\mtt{P})|
    =
    |\mtt{A}_1(\mtt{P})|,
    &&|f_I(\mtt{P})|
    =
    |\mtt{A}_3(\mtt{P})|.
    \label{eq:fd=aiP}
  \end{align}
\end{rem}

\begin{rem}
  Since $\Pbot$ (resp. $\Ptop$) is given by the list of $\mtt{v}_{i,j}$
  that $i$ and $j$ change such as Subroutine \ref{algo:fi(p1)} (resp. \ref{algo:fi(p2)}), 
  the output $S_1$ (resp. $S_2$) of Subroutine \ref{algo:fi(p1)} (resp. \ref{algo:fi(p2)}) satisfies
  \begin{align}
    &S_1
    =
    f_I(\Pbot),
    &&S_2
    =
    f_I(\Ptop).
    \label{rem:bot=alg1}
  \end{align}
\end{rem}

The following lemma implies that
for $\bm{x} \in \mathbb{Z}_q^n$, $f_I(\mtt{P})$ (resp. $f_D(\mtt{P})$) is
a set of insertion (resp. deletion) indices of $\bm{y} \in \mathbb{Z}_q^m$.

\begin{lem}\label{lem:dpin_kai}
  For any $\bm{x} \in \mathbb{Z}_q^n$ and $\bm{y} \in \mathbb{Z}_q^m$,
  any $\mtt{P} \in \Pset$ satisfies $D_{f_I(\mtt{P})}(\bm{y}) = D_{f_D(\mtt{P})}(\bm{x})$.
\end{lem}

\begin{IEEEproof}
  From Lemma \ref{lem:each_size},
  $|\mtt{A}_2(\mtt{P})| =: b$ is a fixed value
  for any $\mtt{P} \in \Pset$.
  Hence,
  we denote $\mtt{A}_2(\mtt{P})$ by 
  \begin{align}
    \mtt{A}_2(\mtt{P})
    =
    \{
    (\mtt{v}_{i_k,j_k}, \mtt{v}_{i_k-1,j_k-1})
    \mid
    k \in \setint{b}
    \}
    \label{eq:type_beta_P}
  \end{align}
  for some $1 \leq i_1 < i_2 < \cdots < i_b \leq n$ and $1 \leq j_1 < j_2 < \cdots < j_b \leq m$.
  Then,
  from the definition of $\mtt{A}_2$,
  we get $x_{i_k} = y_{j_k}$ for all $k \in \setint{b}$.
  Because $f_D(\mtt{P}) = \setint{n} \setminus \{ i_1, i_2,..., i_{b} \}$,
  \begin{align*}
    D_{f_D(\mtt{P})}(\bm{x})
    &=
    (x_{i_1}, x_{i_2}, \cdots, x_{i_b})
  \end{align*}
  holds.
  Similarly,
  we get
  \begin{align*}
    D_{f_I(\mtt{P})}(\bm{y})
    &=
    (y_{j_1}, y_{j_2}, \cdots, y_{j_b}).
  \end{align*}
  From the above,
  $D_{f_D(\mtt{P})}(\bm{x}) = D_{f_I(\mtt{P})}(\bm{y})$ holds.
\end{IEEEproof}

Fix $\bm{x} \in \mbb{E}_{n,q,t}$ ($t \ge 2$).
Then,
for a given $i \in \setint{n-1}$,
let $\bm{t}^{(i)}$ be the sequence swapped adjacent symbols $x_i, x_{i+1}$,
and
$\mbb{T}$ be the set of $\bm{t}^{(i)}$,
i.e.,
\begin{align*}
  \bm{t}^{(i)}
  &=
  (x_1, ..., x_{i-1}, x_{i+1} x_i, x_{i+2}, ..., x_n),\\
  \mbb{T}
  &=
  \{
  \bm{t}^{(i)}
  \mid
  i \in \setint{n-1}
  \}.
\end{align*}
Define $\mbb{U}^{(i)} := \{ \bm{z} \in \mbb{L}_1(\bm{x}) \mid z_i = z_{i+1} \}$
for $i \in \setint{n-1}$,
and $\mbb{U} := \bigcup_{i \in \setint{n-1}}\mbb{U}^{(i)}$.
In words,
$\mbb{U}$ is the set of sequences $\bm{z} \in \mbb{L}_1(\bm{x})$
such that adjacent symbols are the same.
From Lemma \ref{lem:only_exit},
$x_i , x_{i+1}, x_{i+2}$ are distinct for any $i \in \setint{n-2}$.
Hence,
$\mbb{T}$ and $\mbb{U}$ are coprime.

The following holds from the proof of Theorem $1$ of \cite{sala2013counting}
\begin{lem}\label{theo:sala}
  Fix $\bm{x} \in \mbb{E}_{n,q,t}$.
  Define $\mbb{T}$ as above.
  Then,
  $\mbb{D}^{(1)}(\bm{x}) \cap \mbb{D}^{(1)}(\bm{t}^{(i)}) = \{ D_i(\bm{x}), D_{i+1}(\bm{x}) \}$
  for any $i \in \setint{n-1}$.
  If $\bm{y} \in \mbb{L}_1(\bm{x}) \setminus \mbb{T}$,
  $\mbb{D}^{(1)}(\bm{x}) \cap \mbb{D}^{(1)}(\bm{y})$ is a singleton.
\end{lem}

\begin{lem}\label{lem:y_eith}
  Fix $\bm{x} \in \mbb{E}_{n,q,t}$ ($t \ge 2$).
  Define $\mbb{U}^{(i)}$ and $\mbb{U}$ as above.
  For $\bm{y} \in \mbb{U}^{(i)}$,
  $\mbb{D}^{(1)}(\bm{x}) \cap \mbb{D}^{(1)}(\bm{y})$ is a singleton,
  and
  $\bm{z} \in \mbb{D}^{(1)}(\bm{x}) \cap \mbb{D}^{(1)}(\bm{y})$ satisfies
  \begin{align*}
    &\bm{z}
    =
    D_{i}(\bm{y}) = D_{i+1}(\bm{y}),
    &&\bm{z} \neq D_j(\bm{y}),
  \end{align*}
  for $j \in \setint{n-1} \setminus \{i, i+1\}$.
  If $\bm{y} \in \mbb{L}_1(\bm{x}) \setminus \mbb{U}$,
  $\bm{z} \in \mbb{D}^{(1)}(\bm{x}) \cap \mbb{D}^{(1)}(\bm{y})$ satisfies
  the follows for some $i$:
  \begin{align*}
    &\bm{z}
    =
    D_{i}(\bm{y})
    \neq
    D_{j}(\bm{y})
    &&(j \neq i).
  \end{align*}
\end{lem}

We omit the proof by the space \cha{limitations}.

The following lemma means that
the insertion index of $\bm{y} \in \mbb{L}^{(1)}(\bm{x})$ is either $f_I(\Pbot)$ or $f_I(\Ptop)$.

\begin{lem}\label{theo:subset_fip1cfip2}
  For $\bm{x} \in \mbb{E}_{n,q,t}$ with $t \ge 2$ and  $\bm{y} \in \mbb{L}^{(1)}(\bm{x})$,
  define $J := \{ j \in \setint{n} \mid D_{\{j\}}(\bm{y}) \in \mbb{D}^{(1)}(\bm{x}) \}$.
  Then,
  $\Pbot$ and $\Ptop$ given by Definition \ref{def:p_top} satisfy
  \begin{align}
    J
    &=
    f_I(\Pbot) \cup f_I(\Ptop). \label{eq:J_is}
  \end{align}
\end{lem}

\begin{IEEEproof}
  Firstly,
  consider the case for $\bm{y} \in \mbb{L}_1(\bm{x}) \setminus (\mbb{U} \cup \mbb{T})$.
  Lemma \ref{lem:each_size} leads $|f_D(\Pbot)| = |f_I(\Pbot)| = 1$.
  Hence,
  Lemma \ref{lem:dpin_kai} yields
  $D_{f_I(\Pbot)}(\bm{y}) = D_{f_D(\Pbot)}(\bm{x}) \in \mathbb{D}^{(1)}(\bm{x})$.
  Clearly,
  $D_{f_I(\Pbot)}(\bm{y}) \in \mbb{D}^{(1)}(\bm{y})$ holds.
  From the above,
  we have $D_{f_I(\Pbot)}(\bm{y}) \in \mbb{D}^{(1)}(\bm{x}) \cap \mbb{D}^{(1)}(\bm{y})$,
  i.e.,
  $f_I(\Pbot) \subseteq J$.
  Now,
  from Lemmas \ref{theo:sala} and \ref{lem:y_eith},
  $J$ is a singleton
  since $\bm{y} \in \mbb{L}_1(\bm{x}) \setminus (\mbb{U} \cup \mbb{T})$.
  Thus,
  $J = f_I(\Pbot)$ holds
  since $|f_I(\Pbot)|  = 1$.
  Similarly,
  $J = f_I(\Ptop)$ holds.
  Therefore,
  we get \eqref{eq:J_is}.
  
  Secondly,
  consider the case for $\bm{y} \in \mbb{T}$.
  Assume $\bm{y} = \bm{t}^{(i)}$.
  The submatrix of $\mat{H}_{\bm{x}, \bm{y}}$ from rows $i-1$ to $i+2$ and columns $i-1$ to $i+2$
  is given by
  \begin{align*}
    \begin{pmatrix}
      0 ~& 1 ~& 2 ~& 3\\
      1 ~& 2 ~& 1 ~& 2\\
      2 ~& 1 ~& 2 ~& 3\\
      3 ~& 2 ~& 3 ~& 2
    \end{pmatrix}.
  \end{align*}
  Hence,
  we get $f_I(\Pbot) = \{i\}$ and $f_I(\Ptop)=\{i+1\}$.
  Now,
  from Lemmas \ref{theo:sala} and \ref{lem:y_eith},
  $J = \{ i, i+1 \}$ holds
  since $\bm{y} \in \bm{t}^{(i)} \in \mbb{T}$.
  Therefore,
  we get $J = \{ i, i+1 \} = f_I(\Pbot) \cup f_I(\Ptop)$.

  Finally,
  consider the case for $\bm{y} \in \mbb{U}$.
  Assume $\bm{y} \in \mbb{U}^{(i)}$.
  Similarly to the case for $\bm{y} \in \mbb{T}$,
  we get \eqref{eq:J_is}.
\end{IEEEproof}

\subsection{Proof of Theorem \ref{theo:r=m_result_class}}\label{subsec:lem3_proof}
From \eqref{rem:bot=alg1} and Lemma \ref{theo:subset_fip1cfip2},
we get $J = f_I(\Pbot) \cup f_I(\Ptop) = S_1 \cup S_2$.
Now,
$d(\bm{x}, \bm{y})=2$ holds
since $\bm{y} \in \mathbb{L}^{(1)}(\bm{x})$.
Hence,
Lemma \ref{lem:each_size} leads $|\mathtt{A}_3(\Pbot)| = 1$.
Thus,
from \eqref{eq:fd=aiP} and \eqref{rem:bot=alg1},
we have $|S_1| = |f_I(\Pbot)| = |\mathtt{A}_3(\Pbot)| = 1$.
Similarly,
we get $|S_2| = 1$.
Therefore,
$|S_1 \cup S_2| \leq |S_1| + |S_2| = 2$.
\hfill\IEEEQED

\section{Conclusion}
In this paper,
we gave a subroutine making candidates of an insertion index
from a sequence detecting deletion indices
after a single classical insertion and deletion.
By using this,
the received state with single-insertion plus single-deletion is changed
into the state with at most 2 deletions.
By applying a deletion-correcting algorithm to the changed state,
this paper provided a decoding algorithm correcting single-insertion plus single-deletion
for the non-binary deletion-correcting quantum codes constructed by Matsumoto and Hagiwara.

\section*{Acknowledgment}
This work was supported by JSPS KAKENHI Grant Number 22K11905.

\bibliographystyle{IEEEtran}
\bibliography{IEEEabrv,myref}

\end{document}